# ANALYSE CINEMATIQUE DE L'EPAULE CHEZ DES PATIENTS HEMIPARETIQUES LORS DE LA SAISIE D'OBJETS PESANTS

Svetlana **DEDOBBELER**[1], Sylvain **Hanneton**[1], Agnès **Roby-Brami**[1,2]

1 Neurophysique et Physiologie du système moteur, CNRS UMR 8119, 45, rue des Saints-Pères, 75270 Paris
2 Service de Médecine Physique et de Réadaptation, Hôpital Raymond Poincaré, 92380 Garches

Les patients hémiparétiques développent des stratégies motrices de compensation, en particulier une flexion du tronc pour compenser la déficience de la coordination de l'épaule et du coude (Cirstea et Levin, 2000, Roby-Brami et al. 1997, 2003, 2004). Certains patients hémiparétiques sont capables d'effectuer un mouvement coordonné de l'épaule et du coude, mais ils ne le font pas spontanément car ils utilisent de façon automatique des coordinations alternatives de compensation. Les compensations ont une efficacité immédiate sur la capacité fonctionnelle mais pourraient être néfastes à long terme par un mécanisme de learned disuse (Taub et al. 2002).

## *Hypothèse*

Notre hypothèse de travail est que certains patients hémiparétiques sont capables de réapprendre une coordination épaule-coude normale. L'utilisation automatique d'une synergie plus "normale" du membre supérieur pourrait être la base de l'amélioration fonctionnelle dans la vie quotidienne. Afin d'améliorer le mouvement en allant vers le bon geste avant d'utiliser des stratégies de compensation qui sont finalement limitatrices, il est nécessaire de pouvoir le qualifier et le quantifier. Nous avons besoin d'identifier les indices pertinents caractérisant la cinématique du mouvement. Cela permettrait de développer des outils d'aide à la récupération fonctionnelle sur un principe de feedback sensoriel enrichi.

## *Méthodologie*

### *1/La population*

Les patients hémiparétiques sont hospitalisés dans le service de rééducation neurologique de l'hôpital Raymond Poincaré de Garches (Pr. Philippe AZOUVI). Il s'agit de 8 patients adultes, droitiers, de moins de 75 ans, présentant une hémiparésie consécutive à une lésion vasculaire dans la région sylvienne. Les patients recrutés sont capables de donner un consentement éclairé à l'étude avec des troubles associés (aphasie et troubles de la compréhension) faibles à modérés. Autres critères de non inclusion : antécédents d'autres troubles neurologiques ou orthopédiques, maladies intercurrentes, algodystrophie.

### *2/La tâche*

Les sujets sont assis à une table, le tronc est fixé au dossier. Ils réalisent une série de 20 saisies avec le membre atteint puis avec le membre sain. Ils ont pour consigne de prendre l'objet qui leur est présenté et le déposer sur une cible sur la table à distance de bras étendu et surélevée de dix centimètres. Nous leur proposons deux types de poids (200 et 400 grammes), identifiables par la couleur et présentés aléatoirement.

### *3/Le protocole*

Nous utilisons le Polhemus, système électromagnétique de capture du mouvement composé de 4 capteurs (localisés sur l'acromion, la face latérale du bras, la face inféro-postérieure de l'avant bras, et la face dorsale de la main) et d'une source électromagnétique. L'orientation du capteur est définie en terme d'angles d'Euler par l'orientation de son propre repère local par rapport au repère lié à la source émettrice. L'émetteur est à champ étendu d'un rayon de 0.7 m.

Le modèle biomécanique reconstruit est dérivé du modèle développé par Biryukova et al. (2000) et utilisé par Roby-Brami et al. (1997, 2003). Il est visualisé l'aide du logiciel 3D vision (logiciel Kyhopsys).

A travers l'analyse de mouvements capturés puis reconstruits en 3D, nous avons relevé les différences cinématiques ou autres éléments remarquables entre le type de préhension des sujets sains et celui des sujets hémiparétiques, leur réaction au port de charge, leur capacité d'apprentissage et d'automatisation. Nous avons calculé les coordonnées pour le membre sain, le membre paralysé du patient, et le membre dominant des sujets valides. Nous étudions la coordination des différents degrés de liberté, ainsi que les caractéristiques cinématiques du geste.

Dans l'idée d'inciter le patient à « lisser » son mouvement nous relevons en temps réel des indices pertinents : par exemple, la fluidité du mouvement (smoothness), les phases d'arrêts intempestifs ; la quantité de secousses (jerk analysis).

## *Résultats*

Nous avons à disposition toute la cinématique du mouvement ainsi que tous les critères choisis pour des évaluations quantitatives et qualitatives chez des patients hémiparétiques et chez des valides. Ceci nous permettra de faire une analyse complète de l'étude de l'influence du poids sur la saisie d'objet ce qui n'a jamais été fait chez des patients hémiparétiques. Ici nous présentons quelques exemples préliminaires.

*1/Calcul des indices de quantification de la qualité du mouvement*

Les indices sont calculés pour chaque essai (20) chaque sujet, et pour chaque phase (phase 1 : approche, phase 2 : transport et phase 3 : retour). Nous sommes donc en mesure de procéder à des comparaisons des valeurs moyennes des différents indices en considérant
- Les données issues des mouvements accomplis par les patients en comparant le coté atteint au coté « sain »,
- Les données issues des mouvements des personnes « valides » versus les mouvements effectués du coté « sain »,
- Les données issues des mouvements des personnes « valides » versus les mouvements des patients effectués du coté atteint.

*2/Quantification de la qualité du mouvement : quels sont les meilleurs indices ?*

Le premier indice examiné est la moyenne ***des instants où le mouvement atteint sa vitesse maximale***. Nos résultats montrent que cet indice est stable quelque soit la population considérée, que le mouvement concerne le coté sain ou atteint et ceci aussi bien pour la phase 1 (« reaching ») que pour la phase de transport de l'objet (phase 2). L'instant correspondant au pic de vitesse n'est donc pas un bon indice de qualité du mouvement.

La ***vitesse moyenne*** du mouvement semble être un indice plus pertinent. En effet celle-ci est plus faible pour le coté atteint en comparaison au côté sain des patients et la vitesse moyenne des mouvements des personnes valides est plus importante que celle du mouvement des patients (côté sain ou atteint). Ceci est vrai aussi bien pour la phase d'atteinte (phase 1, « reaching ») que pour la phase de transport de l'objet (phase 2). Cependant les différences observées ne sont pas significatives sur l'échantillon considéré. Il est à noter que la vitesse moyenne du mouvement effectué du côté sain des patients n'atteint pas la valeur moyenne obtenue chez les personnes valides.

La ***durée de la phase de mouvement*** est significativement plus importante en moyenne du côté atteint des patients comparée à la durée de cette même phase de mouvement chez les personnes valides. Le mouvement effectué du côté « valide » des patients, dure un peu plus longtemps que chez les personnes valides.

Nous observons dans ces résultats une ***très faible influence du poids de l'objet*** sur les valeurs moyennes des différents indices étudiés et ceci aussi bien pour la phase d'atteinte de l'objet que pour la phase de transport.

*3/Coordination du mouvement : amplitudes articulaires*

Le dispositif expérimental nous permet d'avoir pour chaque essai les données cinématiques ponctuelles (points anatomiques) et angulaires. Une des hypothèses associées à ce travail était que les patients avaient une extension du coude réduite. Nos résultats semblent valider cette hypothèse en particulier lors de la phase 2 (transport et dépôt de l'objet). En effet, l'amplitude moyenne du mouvement d'extension du coude du côté atteint des personnes hémiplégiques est en moyenne inférieure de 20° environ à celle du groupe de personnes valides.

Si l'on regarde l'amplitude de déplacement du « centre » de la scapula dans le plan sagittal, on peut vérifier que les participants respectent la consigne. Les participants handicapés produisent un déplacement significativement plus important que les participants du groupe contrôle.

## *Conclusions*

1/ Deux indices apparaissent comme particulièrement pertinents : le nombre de pics de vitesse et la durée de la phase de mouvement.

2/ La coordination est perturbée chez le patient, il a tendance à avancer avec son tronc pour compenser le développement moindre dans l'espace du membre supérieur. Pourtant les patients ont la capacité fonctionnelle d'utiliser toute l'amplitude mais ne le font pas préférant la compensation du tronc.

3/ Chez le patient, le poids de l'objet a une très faible influence. Ces résultats peuvent être expliqués par le fait que les patients ont une perturbation de la sensibilité et ont des difficultés à ajuster la force de serrage de la main lors de la saisie de l'objet.

Nous sommes particulièrement attentifs à la phase du mouvement de préhension au début du transport où l'influence du poids peut être la plus perturbatrice. Pour aider le patient à fluidifier son mouvement et à corriger sa coordination en limitant les compensations, les indices seront transformés en sons spécifiques dont le patient connaît le sens (la valeur de l'indice influence divers paramètres du signal sonore). Il pourrait alors corriger plus rapidement et lui-même son geste. Ultérieurement, le professionnel pourra choisir le type d'indices à proposer au patient en fonction de sa pathologie.

## *Références*